\documentclass[10pt,english,conference,compsocconf]{IEEEtran}
\usepackage[T1]{fontenc}
\usepackage[latin9]{inputenc}
\usepackage{graphicx}

\makeatletter
\ifCLASSINFOpdf
\else
\fi

\IEEEoverridecommandlockouts

\makeatother

\usepackage{babel}
\begin{document}

\title{Assessment Model for Opportunistic Routing\vspace{-0.7 cm}}

\author{\IEEEauthorblockN{Waldir Moreira, and Paulo Mendes} \IEEEauthorblockA{SITI{*},
University Lusófona, Lisbon, Portugal\\
 Email: \{waldir.junior, paulo.mendes\}@ulusofona.pt} \and\IEEEauthorblockN{Susana
Sargento} \IEEEauthorblockA{IT, University of Aveiro, Aveiro, Portugal\\
 Email: susana@ua.pt} \vspace{-2 cm}\thanks{''This is the author's
preprint version. Personal use of this material is permitted. However,
permission to reprint/republish this material for advertising or promotion
or for creating new collective works for resale or for redistribution
to thirds must be obtained from the copyright owner. The camera-ready
version of this work has been published at LatinCom'11, date of October
2011 and is property of IEEE.''} }
\maketitle
\begin{abstract}
Due to the increased capabilities of mobile devices and through wireless
opportunistic contacts, users can experience new ways to share and
retrieve content anywhere and anytime, even in the presence of link
intermittency. Due to the significant number of available routing
solutions, it is difficult to understand which one has the best performance,
since all of them follow a different evaluation method. This paper
proposes an assessment model, based on a new taxonomy, which comprises
an evaluation guideline with performance metrics and experimental
setup to aid designers in evaluating solutions through fair comparisons.
Simulation results based on the proposed model revisit the performance
results published by \emph{Epidemic}, \emph{PROPHET}, and \emph{BubbleRap},
showing how they perform under the same set of metrics and scenario. \end{abstract}

\begin{IEEEkeywords}
opportunistic routing; assessment model\vspace{-0.1 cm}
\end{IEEEkeywords}

\section{Introduction\vspace{-0.1 cm}}

The increasing capability of portable devices allow users to quickly
form networks by sharing resources (i.e., processing, storage) and
connectivity. Such spontaneous networks are possible by taking advantage
of opportunistic contacts among nodes that can carry and forward information
on behalf of other nodes, allowing information to reach a given destination
even in the presence of intermittent connectivity (resulting from
node mobility, power-saving schemes, physical obstacles).

Several routing proposals emerged taking advantage of devices capability
to overcome intermittency. In such proposals, devices process and
store data until another good intermediate node or destination is
found, based on the store-carry-and-forward paradigm. Routing proposals
range from approaches using node mobility to flood the network for
fast delivery, up to approaches able of controlling such flooding
based on encounter history, prioritization, and encounter prediction.
In recent years, approaches based on social similarity metrics emerged
making use of social relationships, interests, and popularity to improve
opportunistic routing.

This paper starts by proposing an assessment model, which we call
Universal Evaluation Framework (UEF), comprising a set of parameters
related to network density and traffic aiming to aid designers to
carry out fair comparisons among proposals and effectively assess
their performance. It is based on a classification model which identifies
common properties (i.e., routing strategy and metrics) among opportunistic
routing prior-art, aiming to support an efficient development of new
proposals. Then, we compare the performance of \emph{Epidemic} \cite{epidemic},
\emph{PROPHET} \cite{prophet}, and \emph{BubbleRap }\cite{bubblerap}\emph{
}under the conditions specified in the UEF.

This paper is organized as follows. Section \ref{sec:Related-Work}
analyzes work related to existing routing strategy classifications
and evaluation models. The new classification and UEF are presented
in Sections \ref{sec:Taxonomy-for-Opportunistic} and \ref{sec:Evaluation-Model},
respectively. In Section \ref{sec:Fair-Evaluation}, results from
a fair assessment between \emph{Epidemic}, \emph{PROPHET}, and \emph{BubbleRap}
based on the proposed assessment model are discussed. Finally, Section
\ref{sec:Summary-and-Conclusions} concludes the paper.\vspace{-0.15 cm}

\section{Related Work \label{sec:Related-Work}\vspace{-0.15 cm}}

Our motivation to propose an assessment model comes from the fact
that opportunistic routing proposals do not consider neither a similar
set of performance metrics nor comparable experimental scenarios.
This results in unfair comparisons between proposals, since evaluation
metrics and conditions may vary. In this section, we present different
classifications of routing strategies that helped us achieve our goal
of proposing a UEF.\vspace{-0.19 cm}

\subsection{Classification of Routing Strategies\label{sub:Opportunistic-Routing-Strategies}}

Different proposals found in the literature classify routing strategies,
but few provide a way to somehow evaluate them.

Jain et al. (2004) \cite{oracleknowledge} classify proposals based
on knowledge from network oracles or route computation and determination.
A routing evaluation framework is provided considering the different
knowledge oracles to evaluate the amount of knowledge each proposal
requires in specific application scenarios.

Zhang (2006) \cite{zhang06} provides a classification with two main
categories, deterministic and stochastic, which also considers the
type of knowledge used for routing. The main goal is to solely categorize
routing solutions based on the information used to perform data exchange.

Likewise Jain et al. and Zhang, D'Souza and Jose (2010) \cite{survey2010}
classify solutions considering the required knowledge. Proposals are
divided into three major categories based on flooding, history, and,
special devices. It is important to note that this taxonomy succeeds
in including the new social-aware routing trend observed in the last
three years.

The most recent classification proposed by Spyropoulos et al. (2010)
\cite{tax2010} groups opportunistic routing proposals according to
the message exchange scheme they employ: forwarding, replication,
and coding. Authors also identify different types of utility functions
that can be applied to such schemes. Additionally, Disruption Tolerant
Networks (DTN) are classified based on characteristics that have major
impact on routing such as connectivity and mobility. Authors succeed
in mapping routing solutions to the different DTN types, which allows
them to evaluate proposals accordingly.

We observe that the main goal of these classifications is uniquely
to identify the different families of routing solutions. Some of them
\cite{oracleknowledge,tax2010} provide evaluation principles that
simply aid designers to identify the application requirements in order
to propose the ``right'' algorithm. We, on the other hand, use our
classification to identify common aspects among the analysed solutions
to propose a fair way to assess routing performance independently
of the amount of needed knowledge and application scenario.\vspace{-0.15 cm}

\subsection{Evaluation models\vspace{-0.15 cm}}

Regarding evaluation models, we highlight a proposal based on \emph{Evolving
Graph} theory to design and evaluate least cost routing protocols.
Ferreira et al. (2010) \cite{eg} use a formalized metric (i.e., foremost)
to determine journeys (i.e., future temporary connections between
nodes that may provide a path over time) that quickly reach the destination. 

This model provides an algorithm that has good performance when connectivity
patterns are known, and is used as a lower bound reference to compare
MANET routing solutions. Still, it lacks a guideline of how performance
metrics and experimental setups can be used. We, on the other hand,
want to identify the different families of routing solutions according
to their distinct goals and routing strategies, and to provide a set
of experimental setups to aid in a fair performance assessing of opportunistic
routing solutions. \vspace{-0.1 cm}

\section{Classification Model \label{sec:Taxonomy-for-Opportunistic}\vspace{-0.1 cm}}

Observing the opportunistic routing approaches present in the literature,
it is clear the existence of different trends based on specific goals.
Thus, there is the need for a well-balanced and updated taxonomy able
to include trends identified over the last years, focusing solely
on the branch that best reflects opportunistic contacts (i.e., stochastic
\cite{zhang06}). In this section, we briefly describe this new taxonomy
and direct the reader to a detailed version in our report \cite{survey}.

\begin{figure}
\begin{centering}
\includegraphics[scale=0.41]{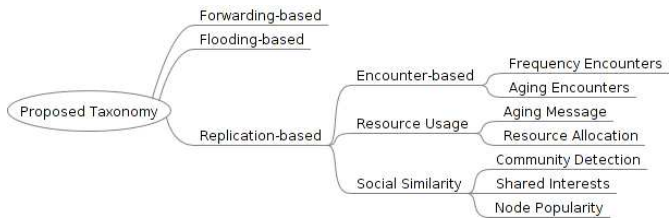}\vspace{-0.2 cm}
\par\end{centering}

\protect\caption{\label{fig:myTaxonomy-1}Taxonomy for opportunistic routing.}
\vspace{-0.6 cm}
\end{figure}

The taxonomy in Fig.~\ref{fig:myTaxonomy-1} considers the analysis
of proposals (spanning a ten-year period) and evolved from a previous
analysis done in \cite{crc}. It starts by identifying three major
categories based on \emph{forwarding}, \emph{flooding}, and \emph{replication}. 

The \emph{forwarding-based} category, also known as \emph{single-copy
forwarding}, is quite interesting from the resource consumption viewpoint,
as proposals are able to spare network and node resources. However,
they suffer from high delay rates, which results in low delivery rate.
Examples of this category are presented by Spyropoulos et al. (2008)
\cite{single-copy-fwd1}.

Albeit being rather aggressive, \emph{flooding-based} algorithms are
capable of achieving high delivery rates, but with a high cost (i.e.,
resource consumption). The classical example of this category is the
\emph{Epidemic} \cite{epidemic} approach. 

In order to reduce resource waste, \emph{replication-based} proposals
aim at increasing delivery rate by adding extra message copies in
the network in a controllable manner. Due to the different routing
algorithms and metrics, \emph{replication}-\emph{based} solutions
are further divided based on encounters, resource usage or social
similarities.

In the \emph{encounter}-\emph{based }category, proposals consider
the history of encounters with a specific destination to support opportunistic
forwarding of messages. The frequency nodes met in the past, or the
time elapsed since the last encounter with the destination is used
to decide about next hops. \emph{PROPHET} \cite{prophet} falls into
this sub-category.

Proposals based on \emph{resource usage} aim to avoid messages to
be kept being forwarded in the network, occupying resources, by creating
metrics that define the age of message copies. They also take forwarding
decisions that wisely use available resources (e.g., \emph{RAPID}
\cite{rapid}). 

In the \emph{social similarity }category, proposals follow more complex
algorithms aiming at: i) avoiding flooding with high probability;
and, ii) exploiting social behavior related to community detection,
shared interest, and node popularity. \textit{Bubble Rap} \cite{bubblerap}
is a good representative of this category.\vspace{-0.05 cm}

\section{Universal Evaluation Framework\label{sec:Evaluation-Model}\vspace{-0.05 cm}}

Our study shows that evaluation methods used so far do not always
consider a homogeneous set of parameters or comparable experimental
setups, which endangers the veracity and fairness of conclusions.
Thus, even with a stable taxonomy, there is the need to devise an
evaluation model based on common performance metrics and experimental
scenarios, avoiding future assessments considering irrelevant performance
metrics and bias scenarios.

To achieve such goal, we looked at seventeen proposals (2000/2010)
that best represented the categories in our taxonomy to identify evaluation/comparison
patterns. For a more detailed view of our findings, refer to our report
\cite{survey}.\vspace{-0.1 cm}

\subsection{Performance Metrics}

Previously \cite{crc}, we checked several metrics and observed the
lack of a naming convention and the use of the same performance metric
with different definitions. This different understanding surely influences
performance assessment.

We observe that delivery rate, cost, and delay are the most used metrics.
Thus, to establish a naming convention and homogeneous definition,
these metrics are defined next. 

Delivery rate is defined as \emph{the number of messages delivered,
per unit of time, out of the total number of messages created}. This
is an important metric, because it reflects the proposal effectiveness.
Proposals able to reach high delivery rates are classified as having
good performance. However, there is always a cost associated to the
delivery of each message. We define cost as \emph{the number of replicas
per delivered message}. This number of replicas gives an idea of how
much resources are consumed. Additionally, message utility is correlated
to its TTL, and it is imperative to remove stale data from the network
to avoid resource waste. Hence, it is important that messages arrive
to their destination within useful time. So, we define delay as \emph{the
time required to deliver all the bytes encompassing a message}.\vspace{-0.2 cm}

\subsection{Experimental Scenario\vspace{-0.1 cm}}

Another concern is the experimental setup in which no rules are followed
as parameters vary in evaluations found in the literature. We observe
that the most common performance assessment parameters considered
among the seventeen analyzed proposals were the number of nodes and
source-destination pairs, meeting times and time between meetings,
area size, message size, network load, message TTL, buffer size, mobility
model, node speed, transmission range, and beacon usage.

We could observe that proposals really differ regarding the experimental
scenario. Some proposals provide detailed information while others
provide it only partially. We were also able to identify two main
parameter classes: network density (network area, number of nodes,
mobility model, transmission range and beacon control), and traffic
(distribution of sources/destinations, load generation, message size,
message TTL, and buffer size). Tables \ref{tab:Network-density-parameters.}
and \ref{tab:Traffic-parameters.} show what UEF recommends along
with what is used in \emph{PROPHET} and \emph{BubbleRap}. We limit
our discussion to these proposals since both are benchmarks for comparison
studies. The former is under standardization process in the DTN research
group and best represents the encounter-based category, and the latter
is a good example of a new trend based on social similarity. Thus,
we start by giving an overview of what the UEF suggests for each main
parameter class and highlight what each proposal considered in their
assessment study.

Network density allows an understanding about behavior on sparse (sporadic
contacts, observing delivery rate when delay is high) and dense (frequent
contacts, assessing the ability to cope with randomness when choosing
next hops) scenarios. Such sparseness may be tuned by configuring
the number of nodes. In average, the number of nodes in the seventeen
analyzed proposals lie between 100-150 (excluding extreme cases as
\emph{FRESH} \cite{fresh}), which are the values proposed in UEF
to define network density. 

In addition, mobility models should consider different speed and pause
time as nodes represent people and vehicles. These parameters do influence
contact and inter-contact times. We observe that some approaches considered
these times as exponentially and power law distributions \cite{rapid,bubblerap},
whereas others obtained them from datasets.

We also analyze the transmission range and beacon usage. In what concerns
the former, we propose ranges from 10-250 meters, since they should
represent the devices capabilities. Yet, beacons should not be used
very often, as battery needs to be spared. However, using it rarely
may lead to losing good contact opportunities. Thus, beaconing at
every 100 ms may provide sufficient network knowledge while saving
energy. We point out that the usage of this parameter in the UEF requires
further investigation to be validated.

Concerning network density parameters (cf. Table \ref{tab:Network-density-parameters.}),
\emph{PROPHET} considers the same number of nodes (50) for different
area densities and mobility models. This is a very interesting approach
as the proposal is subject to scenarios with sporadic contacts/long
delays (Random Waypoint) and frequent contacts/many forwarding opportunities
(Community). Another good point is that this proposal is evaluated
under a mobility model that attempts to mimic human behavior. It is
imperative to consider transmission ranges that actually represent
the different devices capabilities, and the authors also evaluated
the proposal under different ranges. \vspace{-0.2 cm}

\begin{table}[h]
\protect\caption{\label{tab:Network-density-parameters.}Network density parameters}
\vspace{-0.2 cm}

\centering{}\includegraphics[scale=0.42]{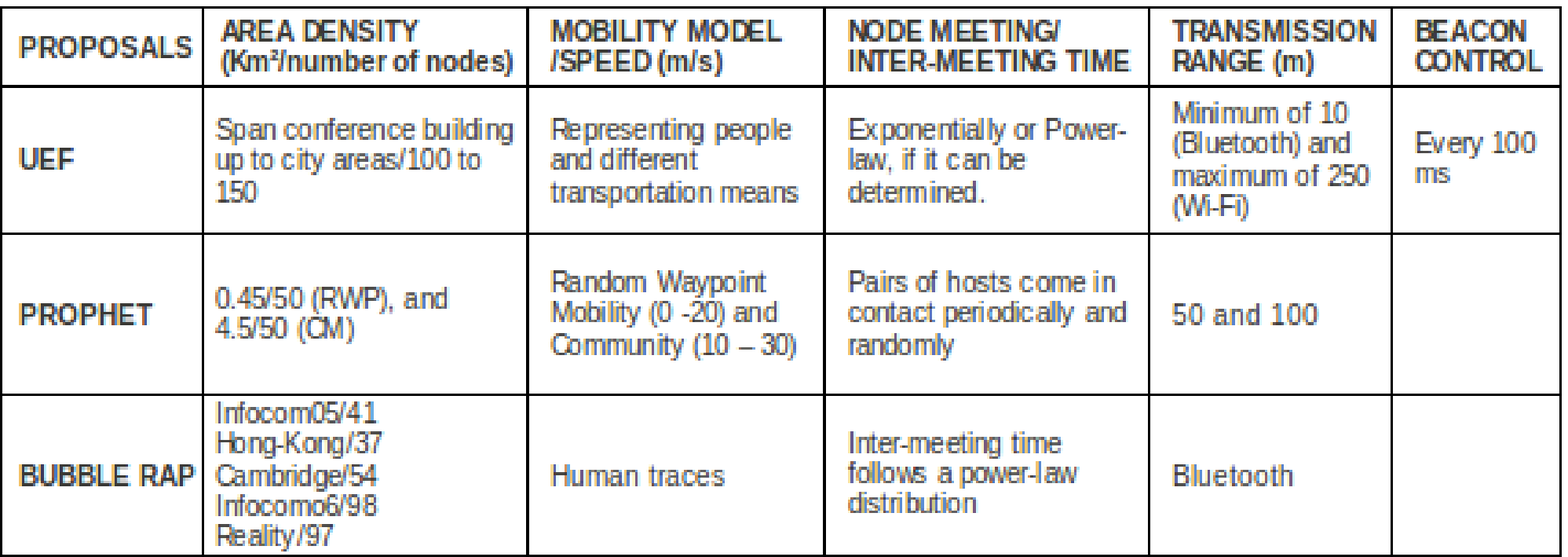}\vspace{-0.2 cm}
\end{table}

Despite considering different area densities and mobility models,
the evaluation of \emph{PROPHET} is done in homogeneous scenarios.
This is not realistic as there are nodes moving according to different
patterns which suggests that, for a better performance evaluation,
the scenario must consider different mobility models simultaneously.
Additionally, node speed must also comply with reality and should
be specified according to whomever/whatever is carrying the device.
The same applies to transmission range that should suitably represent
the capabilities of devices: 10-250 meters shall approximate the simulation
to the real world as they can represent Bluetooth and 802.11b cards,
for instance.

As for \emph{BubbleRap}, since its evaluation is fully based on human
traces, authors succeed in having different area densities (from conference
to city-wide areas). Although no mobility model is explicitly used,
\emph{BubbleRap} follows the human behavior found in traces, which
also results in appropriate node meeting/inter-meeting times \cite{rapid,bubblerap}.
The transmission range considered was that of Bluetooth, which represents
the devices used for collecting the traces. 

When compared to \emph{PROPHET}, \emph{BubbleRap} stands out in terms
of acceptable UEF network density parameters. However, its evaluation
is done in a static manner, i.e., communities are formed and betweenness
centrality is determined based on collected information. Our interest
is to see how the proposal behaves in a dynamic scenario, which shall
influence the way centrality and communities are computed.

Regarding traffic, we observe that, at almost every proposal, the
number of source-destination pairs was statically defined and randomly
assigned. We see no problem in having this number statically defined,
but we certainly agree that changing it during the same experiment
or in different ones shall impose some challenges. However, this number
should be the same and, most importantly, the pairs should remain
the same (which cannot be assured with random choice) to guarantee
similar conditions for assessment study. 

Load generation is a parameter that adds more variations to experiments.
We observe proposals generating a message per second \cite{epidemic,prophet},
a number of messages uniformly distributed \cite{bubblerap} as well
as providing little to no information on the load used. We believe
it must be carefully addressed and homogenized to guarantee fair comparisons.

As applications generate different-sized messages, load must be considered
as it reflects network and node resource consumption. It should be
tuned to represent the different applications that are expected in
a opportunistic scenario (e.g., chat messages, email, file transfer).
We observe that only few proposals provide information about this
parameter (1 KB \cite{epidemic,rapid} or between 10-100 KB \cite{ebr}).

Traffic levels in the network can be affected by message TTL. If the
latter is too high, network and node resource consumption may increase.
Otherwise, messages may not even reach destination. Observed TTL was
defined as the number of hops \cite{epidemic,prophet} or time units,
and normally varied between 3-10 hops in average. Hui et al. \cite{bubblerap}
shows that only 5\% of nodes have some level of relationship with
the destination in the first hop, thus we suggest starting evaluation
with at least 3 hops (as interaction values improve around 35\%),
and varying it to observe the proposals behavior. 

We also observe how a proposal performance can be influenced by buffer
size. This parameter reflects how much of the device storage a user
is willing to sacrifice on behalf of others. Unlimited buffer is not
realistic, whereas providing all space is acceptable in scenarios
where nodes are there to serve others \cite{rapid}. Thus, based on
our observations of the seventeen routing proposals, we suggest to
limit buffer to 200 messages (10KB each).

From Table \ref{tab:Traffic-parameters.} we observe that \emph{PROPHET}
has a different setting for each of the mobility models employed.
Message load varies with the mobility model, and authors evaluate
the proposal considering different message TTL (TTL may vary according
to the message content or application generating it). Normally, buffer
space can be a constraint, since nodes may not be willing to share
it all. So, considering limited buffer space is closer to the real
world scenario, and \emph{PROPHET}'s authors succeed in that.

\begin{table}[h]
\protect\caption{\label{tab:Traffic-parameters.}Traffic parameters}
\vspace{-0.2 cm}

\centering{}\includegraphics[scale=0.42]{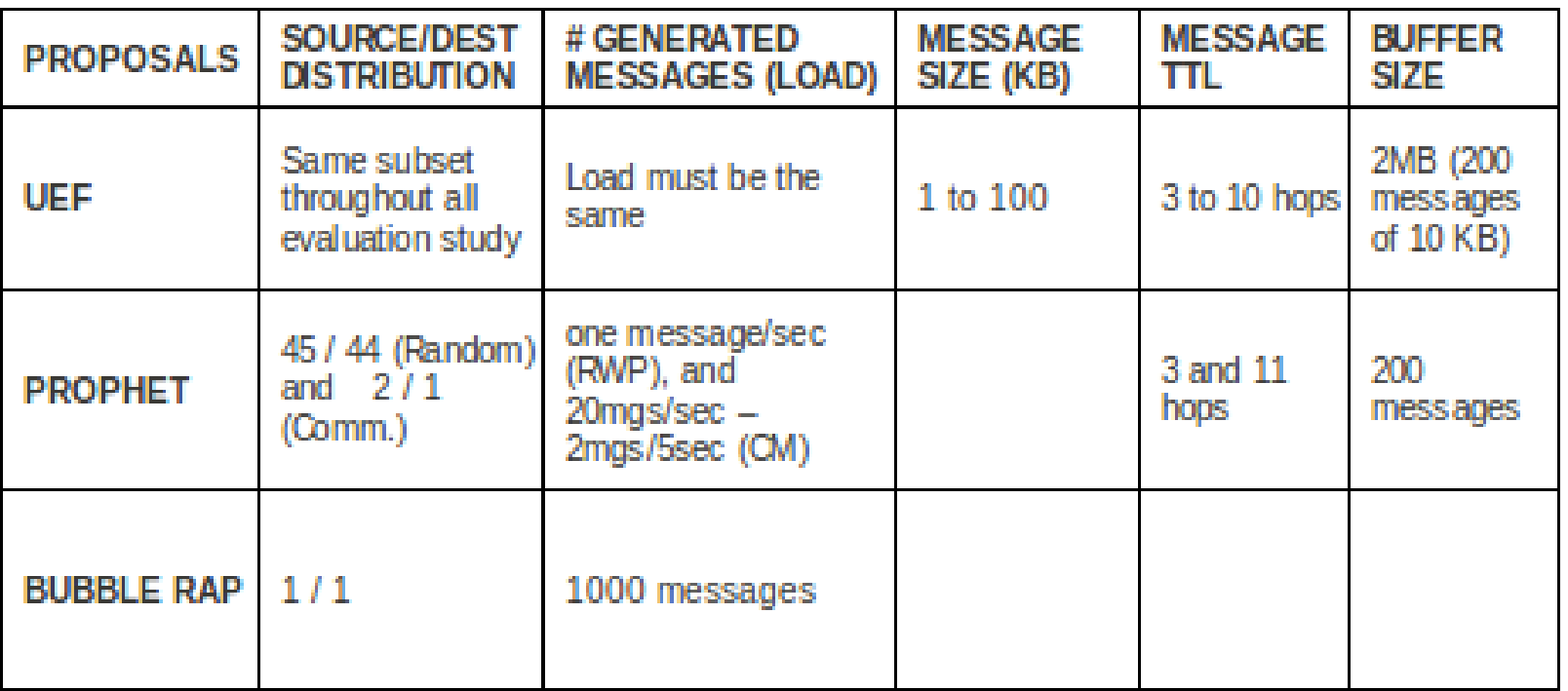}\vspace{-0.6 cm}
\end{table}

Yet, related to these parameters, \emph{BubbleRap} only provides information
on the source/destination distribution and load, where it generates
1000 messages between all node pairs.

This is actually the set of parameters that actually deviates the
most in comparison studies. Normally, source/destination pairs are
a subset of randomly chosen nodes. This influences the evaluation
assessment, as this set will vary as simulations are run. Thus, evaluation
studies should consider the same subset of source/destination pairs
and the number of generated messages must also remain the same.

In both proposals, authors do not mention anything about the message
size. This is an important parameter as real world applications generate
messages with different sizes. Despite having the concern of sparing
buffer, \emph{BubbleRap} does not indicate what size was considered
for evaluation.

Although not explicitly stated, \emph{BubbleRap} considers message
TTL and suggests that a minimum of no less than 3, since most of the
nodes first met (1 and 2 hops) still belong to the community of the
message's carrier. Thus, in order for the message to reach nodes from
the same community as the message's destination, more hops should
be considered. \vspace{-0.5 cm}

\begin{figure*}[t]
\centering{}\includegraphics[scale=0.35]{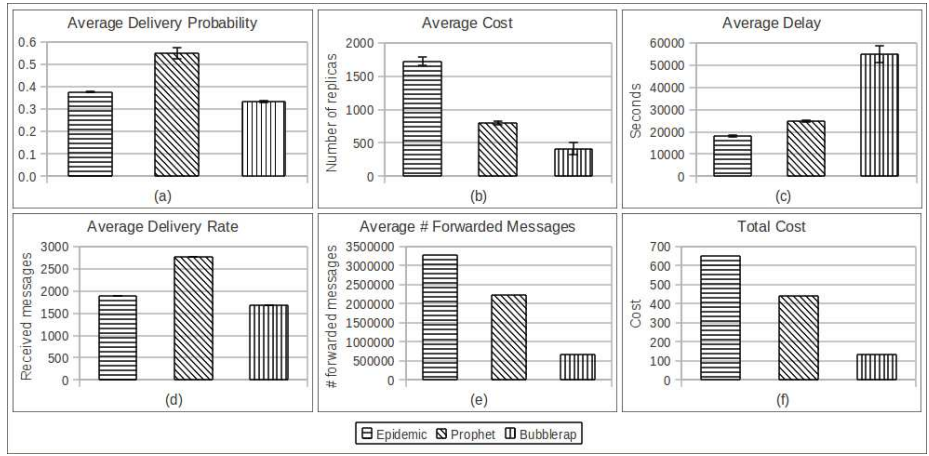}\vspace{-0.2 cm}\protect\caption{\label{fig:2}Performance evaluation results.}
\vspace{-0.6 cm}
\end{figure*}

\section{Fair Evaluation\label{sec:Fair-Evaluation}\vspace{-0.15 cm}}

In this section we present the performance evaluation of \emph{Epidemic},
\emph{PROPHET}, and \emph{BubbleRap} under the same conditions specified
in the UEF. It is important to note that our goal is twofold: first,
to show the importance of a homogeneous evaluation while assessing
routing proposals from categories as different as flooding-based,
encounter-based, and based on social similarities; and second, to
show how the usage of different parameter setups and performance metrics
can result in assessment that is bias to some proposals. \vspace{-0.2 cm}

\subsection{Evaluation settings and methodology\vspace{-0.15 cm}}

Following the proposed UEF, the considered scenario has 150 nodes
distributed in 17 groups (8 of people and 9 of vehicles). One of the
vehicle groups, with 10 nodes, follows the Shortest Path Map Based
Movement mobility model \cite{one} and represent police motorbike
patrols. They move with speed between 7-10 m/s and have a waiting
time between 100-300 seconds when arriving at the randomly chosen
destination.

The other vehicle groups represent buses routes. Each group is composed
of 2 vehicles. They follow the Bus Movement mobility model \cite{one}
with speeds between 7-10 m/s and have waiting times between 10-30
seconds.

Regarding people groups, they follow the Working Day Movement mobility
model \cite{one} with walking speeds ranging from 0.8-1.4 m/s. People
may also use buses to move around. Each of these groups have different
meeting spots, as well as offices and home locations. People spend
8 hours at work and present 50\% probability of having an evening
activity after leaving work. In the office, nodes move around and
have a pause time ranging from 1 minute to 4 hours. Evening activities
can be done alone or in groups with a maximum of 3 people, and can
last between 1 to 2 hours. 

Every node is equipped with a wireless interface with transmission
speed of 11 Mbps and range of 100 meters. 

Traffic load is previously configured with established source/destination
pairs, where approximately 500 messages are generated per day among
the same subset of node pairs. Message size ranges from 1 to 100 kB
and TTL is set at 24 hours. The buffer space is of 2 MB. These values
comply with different applications, and the assumption about user's
limited willingness to share storage capacity. 

Simulations are run on Opportunistic Network Environment \cite{one}
and represent a 12-day interaction between nodes (with 2 days of warmup,
which is not considered for the results). Each simulation is run ten
times (with different random number generator seeds for the used movement
models) in order to provide a 95\% confidence interval for the results.
All the results are evaluated considering the average delivery probability,
cost, and delay.\vspace{-0.1 cm}

\subsection{Results\vspace{-0.1 cm}}

Fig. \ref{fig:2} shows the performance metrics considered in the
UEF (a, b, and c), as well as the ones used in the evaluation of \emph{PROPHET}
(c, d, and e) and \emph{BubbleRap} (a and f).

When comparing Epidemic and \emph{PROPHET} under the UEF scenario,
it is still observed the same behavior reported by Lindgren et al.
(2003) \cite{prophet}: \emph{PROPHET} has better performance by delivering
over 860 more messages (Fig. \ref{fig:2}d) with 32\% less forwardings
(Fig. \ref{fig:2}e) than Epidemic. However, \emph{PROPHET} average
delay (Fig. \ref{fig:2}c) is much higher (6661.3s) than the one of
\emph{Epidemic}, presenting a different behavior than the one reported
by Lindgren et al., where \emph{PROPHET} is able to deliver messages
in less time.

Generally speaking, the performance of \emph{PROPHET} in the UEF scenario
presents higher average delivery rate, forwarded messages, and average
delay than the one reported in the original paper. It needed (roughly
and based on the results presented in its original paper) far more
forwardings (over 340 times) to deliver almost 2.6 more messages with
a delay close to 10 times more. Regarding the UEF performance metrics,
we still believe that the overhead (i.e., average cost) is quite too
exhaustive for a small increase in delivery probability. The increase
in cost (Fig. \ref{fig:2}b) is explained by the message TTL (set
at 24 hours), which allows messages to propagate (i.e., replicate)
more in the system. This consequently contributes to the increase
in the average delay (Fig. \ref{fig:2}c) as messages are held for
a wiser decision. In addition, it is expected that the number of delivered
messages also increase due to messages being further replicated in
the system.

The results of \emph{PROPHET }with the UEF show an example of a proposal
that was shown to have suitable performance in the original paper,
but end up with not so interesting results (e.g., more overhead) in
more heterogeneous scenarios.

Now, moving on to the performance comparison between \emph{PROPHET}
and \emph{BubbleRap} under the UEF scenario. Here we consider the
delivery success ratio (a.k.a., average delivery probability) and
total cost as presented by Hui et al. (2008) \cite{bubblerap}. We
observe the same performance behavior, but like it happened between
\emph{Epidemic} and \emph{PROPHET}, the difference in the gains are
much more evident.

Regarding the delivery success ratio (Fig. \ref{fig:2}a), the difference
is over 17 percentage points between these proposals for a 24-hour
TTL (i.e., 1 day), whereas such difference is reported to be very
subtle in \emph{BubbleRap} original paper. As for the total cost,
the gap is even bigger, going from \textasciitilde{}40\% in the original
paper to \textasciitilde{}70\% with UEF. Thus, in terms of cost, \emph{BubbleRap}
has a very interesting behavior, but with a lower delivery ratio than
reported in the original paper.

It is also observed that while the delivery success ratio of both
solutions lay around 15\% for a 24-hour TTL in the original paper,
with UEF this success ratio goes up to 55\% and 33\% for \emph{PROPHET}
and BubbleRap, respectively. These results really show the burden
(i.e., time spent) of solutions based on community formation and this
is the main reason for the poor performance of \emph{BubbleRap} with
UEF. Although the protocol is given two days (i.e., warmup period)
so it learns about possible communities, it is still not enough for
the protocol to wisely determine the existing communities.

These results show how important it is to consider a dynamic scenario.
The performance of \emph{BubbleRap} is assessed over static scenarios
(based on human traces), i.e., first communities are formed and centralities
determined, then the proposal uses such information to deliver messages.
Despite considering traces of human interaction, the proposal is not
able to adapt when nodes interact to represent the inherent dynamism.
Consequently, its performance is degraded.

In what concerns the UEF performance metrics, we observe that the
difference in terms of cost (Fig. \ref{fig:2}b) between \emph{Epidemic},
\emph{PROPHET}, and \emph{BubbleRap} is small. This is because UEF
considers the number of replicas proposals require to perform a successful
delivery. We believe this is a reasonable approach since nodes may
be quite resource constrained, so replication decisions must be wisely
taken.\vspace{-0.2 cm}

\section{Conclusions\label{sec:Summary-and-Conclusions}\vspace{-0.2 cm}}

The increased capability of devices allows users to experience new
ways to exchange content through opportunistic contacts. However,
this new form of communication must deal with link intermittency,
which has given rise to different solutions that attempt to lessen
such issue. Nevertheless, it is difficult to understand which one
has the best performance as every one of them has a different evaluation
method.

Analyzing the last ten years, we observed different trends based on
specific goals and with different opportunistic routing solutions.
Thus, we analyzed different proposals according to the identified
trends, collected information on their evaluation process, and found
common properties (i.e., routing strategy and metrics). The result
was a UEF which provides guidelines, based on a taxonomy including
the new trend identified recently (i.e., social similarity), comprising
a set of performance parameters and experimental setup to aid designers
fairly assessing the performance of opportunistic routing solutions.
To validate our principles, we simulate \emph{Epidemic}, \emph{PROPHET},
and \emph{BubbleRap }under the same UEF conditions, and we are able
to see that differences among the proposals are more evident with
UEF.

We believe that we have reached our goal of providing a way for designers
to classify their new solutions as well as fairly assessing them.
As new proposals emerge, we will keep our taxonomy up-to-date with
the latest trends identified in the field of opportunistic networks.\vspace{-0.2 cm}

\section*{Acknowledgment\vspace{-0.2 cm}}

Thanks are due to FCT for financial support via PhD grant (SFRH/BD/62761/2009)
to Waldir Moreira and UCR project (PTDC/EEA-TEL/103637/2008).\vspace{-0.25 cm}\bibliographystyle{ieeetr}
\bibliography{bib-or}
\vspace{-0.2 cm}
\end{document}